\documentclass[aps,preprint,prd,superscriptaddress,showpacs]{revtex4}
\usepackage{epsfig}

\newcommand{\be}{\begin{equation}}
\newcommand{\ee}{\end{equation}}
\newcommand{\ba}{\begin{eqnarray}}
\newcommand{\ea}{\end{eqnarray}}
\newcommand{\rmi}[1]{{\mbox{\scriptsize #1}}}
\newcommand{\fig}{Fig.~}
\newcommand{\eq}{Eq.~}
\newcommand{\se}{Sec.~}
\newcommand{\eqs}{Eqs.~}
\newcommand{\nr}[1]{(\ref{#1})}
\newcommand{\tr}{{\rm Tr\,}}
\newcommand{\nn}{\nonumber \\}
\newcommand{\fr}[2]{{\frac{#1}{#2}\,}}
\newcommand{\msbar}{{\overline{\mbox{\rm MS}}}}
\newcommand{\lambdamsbar}{{\Lambda_{\overline{\rm MS}}}}

\def\lsi{\raise0.3ex\hbox{$<$\kern-0.75em\raise-1.1ex\hbox{$\sim$}}}
\def\gsi{\raise0.3ex\hbox{$>$\kern-0.75em\raise-1.1ex\hbox{$\sim$}}}

\newcommand{\gsim}{\mathop{\gsi}}

\newcommand{\bmu}{\bar\mu}

\newcommand{\aM}{\alpha_\rmi{M}}
\newcommand{\aG}{\alpha_\rmi{G}}
\newcommand{\bM}{\beta_\rmi{M}}
\newcommand{\bG}{\beta_\rmi{G}}
\newcommand{\aE}[1]{\alpha_\rmi{E#1}}
\newcommand{\bE}[1]{\beta_\rmi{E#1}}
\newcommand{\logT}{\ln\frac{\bmu}{4 \pi T}}
\newcommand{\logz}[1]{\frac{\zeta'(-#1)}{\zeta(-#1)}}

\makeatletter \@addtoreset{equation}{section} \makeatother
\renewcommand{\theequation}{\arabic{section}.\arabic{equation}}

\begin{document}


\title{The pressure of hot QCD up to g$^{\,\mbox{\bf\small 6}\,}$ln(1/g)} 

\author{K. Kajantie}

\affiliation{Department of Physics,
P.O.Box 64, FIN-00014 University of Helsinki, Finland}

\author{M. Laine}

\affiliation{Theory Division, CERN, CH-1211 Geneva 23,
Switzerland}

\author{K. Rummukainen}

\affiliation{Department of Physics,
P.O.Box 64, FIN-00014 University of Helsinki, Finland}

\author{Y. Schr\"oder}

\affiliation{Center for Theoretical Physics, 
MIT, Cambridge, MA 02139, USA}

\begin{abstract}
The free energy density, or pressure, of QCD has at high temperatures 
an expansion in the coupling constant $g$, known so far up to 
order $g^5$. We compute here the last contribution which can be 
determined perturbatively, $g^6\ln(1/g)$, by summing together results 
for the 4-loop vacuum energy densities of two different three-dimensional 
effective field theories. We also demonstrate that the inclusion of the new 
perturbative $g^6\ln(1/g)$ terms, once they are summed 
together with the so far unknown 
perturbative and non-perturbative $g^6$ terms, could potentially extend 
the applicability of the coupling constant series 
down to surprisingly low temperatures.
\end{abstract}
 
\pacs{11.10.Wx, 05.70.Ce, 11.15.Bt, 12.38.Bx, 12.38.Mh}

\maketitle

%
\section{Introduction}

Due to asymptotic freedom, the properties of QCD might be expected to 
be perturbatively computable in various ``extreme'' limits, such as high 
virtuality, high baryon density, or high temperature. We concentrate
here on the last of these circumstances, that is 
temperatures~$T$ larger than a few hundred MeV. 

The physics observable we consider is the pressure, 
or minus the free energy density, of the QCD plasma. 
Potential phenomenological applications include the expansion 
rate of the Early Universe after it has settled into the Standard Model 
vacuum, 
as well as the 
properties of the apparently ideal hydrodynamic expansion observed 
in on-going heavy ion collision experiments, 
just shortly after the impact. 

In these environments, 
it turns out that the naive expectation concerning the validity of
perturbation theory is too optimistic. 
Indeed, even assuming an arbitrarily
weak coupling constant $g$, perturbation theory can only be worked
out to a finite order in it, before the serious infrared problems of 
finite temperature field theory deny further
analytic progress~\cite{linde,gpy}. 
For the pressure, the problem is met at the 4-loop 
order, or ${\cal O}(g^6)$. 

This leads to the interesting situation that there is a definite
limit to how far perturbation theory needs to be pushed.  
So far, there are known loop contributions at orders 
${\cal O}(g^2)$~\cite{es},
${\cal O}(g^3)$~\cite{jk},
${\cal O}(g^4\ln(1/g))$~\cite{tt},
${\cal O}(g^4)$~\cite{az}, and
${\cal O}(g^5)$~\cite{zk}. 
There is also an all-orders numerical result available 
for a theory with an 
asymptotically large number of fermion flavors~\cite{gdm}.
The purpose of the present 
paper is to collect together results from two
accompanying papers~\cite{ys,aminusb}, allowing to determine analytically
the last remaining perturbative contribution, ${\cal O}(g^6\ln(1/g))$, 
for the physical QCD. 

It must be understood that even if computed up to such a high order, 
the perturbative expansion could well converge only very 
slowly, requiring perhaps 
something like $T\gg$~TeV, to make any sense 
at all~\cite{zk,bn,adjoint}. With one further coefficient available, 
we can to some extent now reinspect this issue. To do so we actually
also need to assume something about the unknown ${\cal O}(g^6)$ term, 
since the numerical factor inside the logarithm in 
${\cal O}(g^6\ln(1/g))$ remains otherwise undetermined. Therefore, our 
conclusions on this point remain on a conjectural level, 
but turn out to show 
nevertheless a somewhat interesting pattern, which is why we would
like to include them in this presentation. 

Finally, it should be stressed that even if the perturbative expansion 
as such were to remain numerically useless at realistic temperatures, 
these multiloop computations 
are still worthwhile: the infrared problems of finite temperature QCD
can be isolated to a three-dimensional (3d) effective field 
theory~\cite{dr} and studied non-perturbatively there with simple 
lattice simulations~\cite{a0cond}. However, to convert the results 
from 3d lattice regularisation to 3d continuum regularisation, and from 
the 3d continuum theory to the original four-dimensional (4d) physical 
theory, still necessitates a number of perturbative ``matching'' computations.
Both of these steps are very closely related to what we do here, although we 
discuss explicitly only the latter one.

%
\section{The basic setting}

We start by reviewing briefly how it is believed that
the properties of QCD  
at a finite temperature $T$
can be reduced to a number of
perturbatively computable matching coefficients, 
as well as some remaining contributions from a series of
effective field theories~\cite{dr}. Our presentation follows mostly 
that in~\cite{bn}, but there are a few significant differences.  

The underlying theory is finite temperature QCD with the gauge group 
SU($N_c$), and $N_f$ flavors of massless quarks. In dimensional 
regularisation the bare Euclidean
Lagrangian reads, before gauge fixing,   
\ba
 S_\rmi{QCD} & = & \int_0^{\beta\hbar} \! {\rm d}\tau \int \! {\rm d}^d x\, 
 {\cal L}_\rmi{QCD}, \\
 {\cal L}_\rmi{QCD} & = & 
 \fr14 F_{\mu\nu}^a F_{\mu\nu}^a+
 \bar\psi \gamma_\mu D_\mu \psi \;,
\ea
where $\beta = T^{-1}$,  $d=3-2\epsilon$, $\mu,\nu=0,...,d$, 
$F_{\mu\nu}^a = \partial_\mu A_\nu^a - \partial_\nu A_\mu^a + 
g f^{abc} A_\mu^b A_\nu^c$, 
$D_\mu = \partial_\mu - i g A_\mu$, 
$A_\mu = A_\mu^a T^a$, 
$\gamma_\mu^\dagger = \gamma_\mu$, 
$\{\gamma_\mu,\gamma_\nu\} = 2 \delta_{\mu\nu}$, 
and $\psi$ carries Dirac, color, and flavor indices.   

Denoting the generators 
of the adjoint representation by $(F^a)_{bc} = - i f^{abc}$, 
we define the usual group theory factors, 
\ba
 C_A \delta_{ab} & = & [F^c F^c]_{ab} \;, \hspace*{2.5cm}
 C_F \delta_{ij}\; =\; [T^a T^a]_{ij} \;, \\
 T_A \delta^{ab} & = & \tr F^a F^b    \;, \hspace*{2.5cm}
 T_F \delta^{ab}\; =\; \tr T^a T^b    \;, \\
 d_A & = & \delta^{aa} = N_c^2 - 1 \;, \hspace*{2.2cm}
 d_F\; =\; \delta_{ii} = T_F d_A/C_F \;.
\ea
Obviously $T_A = C_A$. For the standard normalisation, with $N_f$
quark flavors, $C_A = N_c, C_F = (N_c^2 - 1)/(2 N_c), T_A = N_c, 
T_F = N_f/2, d_A = N_c^2 -1, d_F = N_c N_f$.

We use dimensional regularisation throughout this paper.  
The spatial part of each momentum integration measure is written as 
\be
 \int_p \equiv \int \frac{{\rm d}^dp}{(2\pi)^d} =
 \mu^{-2\epsilon}\biggl[\bmu^{2\epsilon}
 \biggl( \frac{e^\gamma}{4\pi} \biggr)^\epsilon  
 \int \frac{{\rm d}^dp}{(2\pi)^d} \biggr],
\ee
where $\mu = \bmu (e^\gamma/4\pi)^{1/2}$, 
and the expression in square brackets has integer dimensionality.
{}From now on we always assume implicitly that 
the factor $\mu^{-2\epsilon}$ is attached to some relevant coupling 
constant, so that  
the 4d $g^2$ 
is dimensionless,  while the dimensionalities 
of $g_\rmi{E}^2,\lambda_\rmi{E}^{(1)},
\lambda_\rmi{E}^{(2)}$ and $g_\rmi{M}^2$, 
to be introduced presently, are GeV.

The basic quantity of interest to us here is minus
the free energy density $f_\rmi{QCD}(T)$, or the pressure 
$p_\rmi{QCD}(T)$, defined by 
\ba
  p_\rmi{QCD}(T) & \equiv & \lim_{V \to \infty} \frac{T}{V}\ln 
 \int {\cal D}A_\mu^a \, {\cal D}\psi \, {\cal D}\bar\psi
 \exp\Bigl(-{1\over\hbar} S_\rmi{QCD} \Bigr), \label{pqcd} 
\ea
where $V$ denotes the $d$-dimensional volume. Boundary conditions
over the compact time-like direction are periodic for bosons and
anti-periodic for fermions. Moreover,  we assume $p_\rmi{QCD}(T)$ 
renormalised such that it vanishes at $T=0$. To simplify the
notation, we do not show the infinite volume limit explicitly
in the following.

At high temperatures and a small coupling, there are parametrically
three different mass scales in the 
problem, $\sim 2 \pi T, gT, g^2T$~\cite{dr}. 
All the effects of the hard mass scale $\sim 2 \pi T$ can be accounted
for by a method called dimensional 
reduction~\cite{dr,generic}. Specifically, 
\ba
 p_\rmi{QCD}(T) & \equiv & p_\rmi{E}(T) 
 + \frac{T}{V}\ln 
 \int {\cal D}A_k^a \, {\cal D}A_0^a  
 \exp\Bigl( - S_\rmi{E} \Bigr), \label{QCD_E} \\
 S_\rmi{E} & = &  \int \! {\rm d}^d x\, {\cal L}_\rmi{E}, \\
 {\cal L}_\rmi{E} & = & \fr12 \tr F_{kl}^2 + \tr [D_k,A_0]^2 + 
 m_\rmi{E}^2\tr A_0^2 +\lambda_\rmi{E}^{(1)} (\tr A_0^2)^2
 +\lambda_\rmi{E}^{(2)} \tr A_0^4 + ...\; . 
 \hspace*{0.5cm} \label{eqcd}
\ea
Here $k=1,...,d$, $F_{kl} = (i/g_\rmi{E}) [D_k,D_l]$, 
$D_k = \partial_k - i g_\rmi{E} A_k$,
and we have used the shorthand notation 
$A_k = A_k^a {\bar T}^a, A_0 = A_0^a {\bar T}^a$, where ${\bar T}^a$
are Hermitean generators of SU($N_c$) normalised such that 
$\tr {\bar T}^a {\bar T}^b = \delta^{ab}/2$. Note that the 
quartic couplings $\lambda_\rmi{E}^{(1)}$, $\lambda_\rmi{E}^{(2)}$
are linearly independent only for $N_c \ge 4$. 

The relation in~\eq\nr{QCD_E} contains five different matching 
coefficients, $p_\rmi{E}, m_\rmi{E}^2, g_\rmi{E}^2, 
\lambda_\rmi{E}^{(1)}, \lambda_\rmi{E}^{(2)}$.
We are interested in the expression for $p_\rmi{QCD}(T)$ up to order 
${\cal O}(g^6 T^4)$. They will then have to be determined to some sufficient
depths, as we will specify later on. Let us here note that the leading
order magnitudes are 
$p_\rmi{E} \sim T^4$, 
$m_\rmi{E}^2 \sim g^2 T^2$, 
$g_\rmi{E}^2 \sim g^2 T$, 
$\lambda_\rmi{E}^{(1)} \sim g^4 T$, 
$\lambda_\rmi{E}^{(2)} \sim g^4 T$.

Apart from the operators shown explicitly in~\eq\nr{eqcd},
there are of course
also higher order ones in ${\cal L}_\rmi{E}$. 
The lowest such operators have been classified in~\cite{sc}. 
Their general structure is that one must add at least two powers 
of $D_k$ or $g A_0$, to the basic structures in~\eq\nr{eqcd}. 
Since higher order operators are generated through interactions
with the scales that have been integrated out, $\sim 2\pi T$, 
they must also contain an explicit factor of at least $g^2$.  
For dimensional reasons, the schematic structure is thus
\be
 \delta {\cal L}_\rmi{E} \sim 
 g^2 \frac{D_k D_l}{(2 \pi T)^2} {\cal L}_\rmi{E} \;. \label{ho_ops}
\ee
To estimate the largest possible contributions such operators
could give, let us assume the most conservative possibility that 
the only dynamical scale in the effective theory is $\sim gT$. 
By dimensional analysis, we then obtain a contribution
\be
 \frac{\delta p_\rmi{QCD}(T)}{T} \sim \delta {\cal L}_\rmi{E} \sim
 g^2 \frac{(gT)^2}{(2 \pi T)^2} (gT)^3 \sim g^7 T^3 \;. 
\ee 
Therefore, all higher dimensional operators can be omitted
from the action in~\eq\nr{eqcd}, if we are only interested 
in computing $p_\rmi{QCD}(T)$ up to order ${\cal O}(g^6 T^4)$. 

The theory in~\eq\nr{eqcd} contains still two dynamical 
scales, $gT, g^2T$.
All the effects of the ``color-electric'' scale, $g T$, can be accounted
for by integrating out $A_0$~\cite{dr}. Specifically, 
\ba
\frac{T}{V}\ln 
 \int {\cal D}A_k^a \, {\cal D}A_0^a  
 \exp\Bigl( - S_\rmi{E} \Bigr) 
 & \equiv & p_\rmi{M}(T) + \frac{T}{V}\ln 
 \int {\cal D}A_k^a 
 \exp\Bigl( - S_\rmi{M} \Bigr), \label{pM} \\
 S_\rmi{M} & = &  \int \! {\rm d}^d x\, {\cal L}_\rmi{M}, \label{SM} \\
 {\cal L}_\rmi{M} & = & \fr12 \tr F_{kl}^2+... \;, \label{mqcd}
\ea
where $F_{kl} = (i/g_\rmi{M}) [D_k,D_l]$, 
$D_k = \partial_k - i g_\rmi{M} A_k$, and 
$A_k = A_k^a {\bar T}^a$. 

The relation in~\eq\nr{pM} contains two
matching coefficients, $p_\rmi{M}, g_\rmi{M}^2$, 
which again have to be determined to sufficient depths. 
At leading order,  
$p_\rmi{M} \sim m_\rmi{E}^3 T$, 
$g_\rmi{M}^2 \sim g_\rmi{E}^2$. 
In addition, there are also higher order operators in~\eq\nr{mqcd}. 
The lowest ones can be obtained by imagining again that we apply
at least two covariant derivatives to~\eq\nr{mqcd}, 
together with at least one factor $g_\rmi{E}^2$ brought in by the 
interactions with the massive modes. This leads to an operator
\be
 \delta {\cal L}_\rmi{M} \sim g_\rmi{E}^2 \frac{D_k D_l}{m_\rmi{E}^3} 
 {\cal L}_\rmi{M} \;.
\ee 
The only dynamical scale in the effective theory being $\sim g^2T$,  
dimensional analysis indicates that we then obtain a contribution
of the order
\be
 \frac{\delta p_\rmi{QCD}(T)}{T} \sim \delta {\cal L}_\rmi{M} \sim 
 g_\rmi{E}^2 \frac{(g^2 T)^2}{m_\rmi{E}^3} (g^2 T)^3 \sim g^{9} T^3 \;. 
\ee 
Therefore, higher dimensional operators can again be omitted, 
if we are only interested in the order ${\cal O}(g^6 T^4)$ for
$p_\rmi{QCD}(T)$. 

After the two reduction steps, there still remains 
a contribution from the scale $g^2 T$, 
\ba
 p_\rmi{G}(T) & \equiv & \frac{T}{V}\ln 
 \int {\cal D}A_k^a 
 \exp\Bigl( - S_\rmi{M} \Bigr) \;, \label{gqcd}
\ea
with $S_\rmi{M}$ in~\eqs\nr{SM}, \nr{mqcd}. Since ${\cal L}_\rmi{M}$ 
only has one parameter, and it is dimensionful, the 
contribution is of the form
\be
 p_\rmi{G}(T) \sim T g_\rmi{M}^6. 
\ee
The coefficient of this contribution is, however, 
non-perturbative~\cite{linde,gpy}. 

In the following sections, we proceed in the opposite
direction with regard to the presentation above, from the ``bottom'' 
scale $g^2 T$, producing $p_\rmi{G}(T)$, through the ``middle'' 
scale $gT$, producing $p_\rmi{M}(T)$, 
back to the ``top'' scale $2\pi T$, producing $p_\rmi{E}(T)$.
We collect on the 
way all contributions up to order $g^6 T^4$,  
to obtain
$p_\rmi{QCD}(T) = p_\rmi{E}(T) + p_\rmi{M}(T) + p_\rmi{G}(T)$.

%
\section{Contributions from the scale $g^2 T$}
\label{se:ggT}

The contribution to $p_\rmi{QCD}(T)$ from the scale $p\sim g^2 T$
is obtained by using the theory ${\cal L}_\rmi{M}$ in~\eq\nr{mqcd}
in order to compute $p_\rmi{G}(T)$, as defined by~\eq\nr{gqcd}. 

As is well known~\cite{linde,gpy}, the computation involves infrared 
divergent integrals, starting at the 4-loop level. This is a reflection 
of the fact that~${\cal L}_\rmi{M}$ defines a confining field theory. 
Therefore, $p_\rmi{G}(T)$ cannot be evaluated in perturbation theory. 

What can be evaluated, however, is the logarithmic ultraviolet divergence 
contained in $p_\rmi{G}(T)$. For dimensional reasons, the  
non-perturbative answer would have to be of the form
\be
 \frac{p_\rmi{G}(T)}{T \mu^{-2 \epsilon} } = 
 d_A C_A^3 \frac{g_\rmi{M}^6}{(4\pi)^4} 
 \biggl[ \aG \Bigl( 
 \frac{1}{\epsilon} + 8 \ln\frac{\bmu}{2 m_\rmi{M}}
 \Bigr) + \bG + {\cal O}(\epsilon)   
 \biggr]
 \;, \label{resG}
\ee
where $m_\rmi{M} \equiv C_A g_\rmi{M}^2$.
Now, because of the super-renormalisability of ${\cal L}_\rmi{M}$, 
the coefficient $\aG$ can be computed in 4-loop perturbation theory, 
even if the constant part $\bG$ cannot~\footnote{%
   The constant part $\bG$
   could be determined by measuring a suitable 
   observable on the lattice and converting then the 
   result from lattice regularisation to the $\msbar$ scheme 
   by a perturbative 4-loop matching computation.}. 

Of course, if we just carry out the 4-loop
computation in strict dimensional regularisation, then
the result vanishes, because there are no perturbative mass scales 
in the problem. This means that ultraviolet and infrared divergences
(erroneously) cancel against each other. 
Therefore, we have to be more careful in order to determine $\aG$. 

To regulate the infrared divergences, we introduce by hand a mass scale, 
$m_\rmi{G}^2$, into the gauge field (and ghost)
propagators. This computation is 
described in detail in~\cite{ys}. Individual diagrams contain then
higher order poles, like $1/\epsilon^2$, as well as a polynomial
of degree up to nine in the gauge parameter $\xi$. However, terms
of both of these types cancel in the final result, 
which serves as a nice check of the procedure. 

As a result, we obtain
\be 
 \frac{p_\rmi{G}(T)}{T \mu^{-2 \epsilon}  }  \approx 
 d_A C_A^3 \frac{g_\rmi{M}^6}{(4\pi)^4} 
 \biggl[ \aG \Bigl( 
 \frac{1}{\epsilon} + 8 \ln\frac{\bmu}{2 m_\rmi{G}}
 \Bigr) + \tilde \bG (\xi) + {\cal O}(\epsilon)   
 \biggr] \;,
\ee
where ``$\approx$'' is used to denote that 
only the coefficient $\aG$ multiplying $1/\epsilon$
is physically meaningful, 
as it contains the desired gauge independent ultraviolet
divergence, defined in~\eq\nr{resG}. The value of the coefficient,
obtained by extensive use of techniques of symbolic computation
(implemented~\cite{ys_proc} in FORM~\cite{jamv}), is~\cite{ys}
\be
 \aG =  
 \frac{43}{96} - \frac{157}{6144} \pi^2 \approx 0.195715 \;. 
 \label{b} \label{aG}
\ee
On the contrary, the constant part  
$\tilde \bG (\xi)$ depends on the gauge parameter $\xi$, because
the introduction of $m_\rmi{G}^2$ breaks gauge invariance, 
and has nothing to do with $\bG$ in~\eq\nr{resG}.

%
\section{Contributions from the scale $gT$}
\label{se:gT}

We next proceed to include the contribution from the scale
$gT$, contained in $p_\rmi{M}(T)$, as defined by~\eq\nr{pM}. 

By construction, \eq\nr{pM} assumes that all the infrared
divergences of the expression on the left-hand-side
are contained in $p_\rmi{G}(T)$, 
defined in~\eq\nr{gqcd}, and 
determined in~\eq\nr{resG}. Therefore, 
if we compute the functional integral
$(T/V)\ln[ \int {\cal D}A_i^a \, {\cal D}A_0^a \exp( - S_\rmi{E})]$
using strict dimensional regularisation (i.e., without 
introducing by hand any 
mass $m_\rmi{G}$ for the gauge field $A_i$), 
whereby $p_\rmi{G}(T)$ vanishes due to the cancellation
between infrared and ultraviolet divergences mentioned above, 
we are guaranteed
to obtain just the infrared insensitive matching coefficient
$p_\rmi{M}(T)$. This is exactly the computation we need, and carry
out in~\cite{aminusb,sd}. It may be mentioned that
we have checked explicitly the 
infrared insensitivity of the result, by giving an equal 
mass to both $A_0$ and $A_i$ in the 4-loop expression for
the functional integral, and then subtracting the graphs responsible
for $p_\rmi{G}(T)$, with the same infrared regularisation. 
This result is also independent of the gauge parameter. 

Keeping terms up to order ${\cal O}(g^6 T^4)$, 
the full outcome for $p_\rmi{M}(T)$ is 
\ba
 \frac{p_\rmi{M}(T)}{T \mu^{-2 \epsilon} } & = &   
 \frac{1}{(4\pi)} 
 d_A  m_\rmi{E}^3 
 \biggl[\fr13 + {\cal O}(\epsilon) \biggr] \nn
 & + & 
 \frac{1 
 }{(4\pi)^2} 
 d_A C_A  
 g_\rmi{E}^2 m_\rmi{E}^2
 \biggl[-\frac{1}{4\epsilon} - \fr34 -\ln\frac{\bmu}{2 m_\rmi{E}} 
 + {\cal O}(\epsilon) \biggr] \nn
 & + & 
 \frac{1 
 }{(4\pi)^3}  
 d_A C_A^2 
 g_\rmi{E}^4 m_\rmi{E}
 \biggl[-\frac{89}{24} - \fr16 \pi^2 + \frac{11}{6}\ln 2  
 + {\cal O}(\epsilon) \biggr] \nn
 & + & 
 \frac{1 
 }{(4\pi)^4}
 d_A C_A^3  
 g_\rmi{E}^6
 \biggl[ \aM \Bigl( 
 \frac{1}{\epsilon} + 8 \ln\frac{\bmu}{2 m_\rmi{E}}
 \Bigr) 
  + \bM + {\cal O}(\epsilon)   
 \biggr] \nn
 & + &  
 \frac{1 
 }{(4\pi)^2} 
 d_A (d_A + 2) 
 \lambda_\rmi{E}^{(1)} m_\rmi{E}^2 
 \biggl[ - \fr14 + {\cal O}(\epsilon)
 \biggr] \nn
 & + &  
 \frac{1 
 }{(4\pi)^2} 
 d_A \frac{2 d_A - 1}{N_c}  
 \lambda_\rmi{E}^{(2)} m_\rmi{E}^2 
 \biggl[ - \fr14 + {\cal O}(\epsilon)
 \biggr] \;, \label{gT_contr}
\ea
where~\cite{aminusb} 
\be
 \aM =  
 \frac{43}{32} - \frac{491}{6144} \pi^2 \approx 0.555017 \;.   \label{aM}
\ee
The finite constant $\bM$ can be expressed in terms of 
a number of finite coefficients related to 4-loop vacuum 
scalar integrals~\cite{aminusb}, 
but we do not need it here. 

In addition to $p_\rmi{M}(T)$, we also need
to specify the effective parameter $g_\rmi{M}^2$ 
appearing in ${\cal L}_\rmi{M}$, to complete contributions from the scale
$g T$. It is of the form 
\be
 g_\rmi{M}^2 = g_\rmi{E}^2 
 \Bigl( 1 + {\cal O}(g_\rmi{E}^2/m_\rmi{E}) \Bigr)  \;, \label{gM}
\ee
where the next-to-leading order correction is 
known (see, e.g.,~\cite{fkrs}), but not needed here. 

%
\section{Contributions from the scale $2 \pi T$}
\label{se:2piT}

The contributions from the scale $2 \pi T$ are contained in 
the expressions for the parameters of the previous effective
theories, as well as in $p_\rmi{E}(T)$.  We write these as 
\ba
 \mu^{2 \epsilon} p_\rmi{E}(T) & = & T^4 \Bigl[\aE{1}
 + g^2 
 \Bigl(\aE{2} + {\cal O}(\epsilon)\Bigr) \nn 
 & & \hspace*{0.7cm}
 + \frac{g^4}{(4\pi)^2} 
 \Bigl(\aE{3} + {\cal O}(\epsilon)\Bigr) 
 + \frac{g^6}{(4\pi)^4} 
 \Bigl(\bE{1} + {\cal O}(\epsilon)\Bigr) + {\cal O}(g^8)
 \Bigr], \label{pE} \\ 
 m_\rmi{E}^2 & = & T^2 \Bigl[ g^2 
 \Bigl( \aE{4} + 
 \aE{5} \epsilon + {\cal O}(\epsilon^2) \Bigr)
 + \frac{g^4}{(4\pi)^2} 
 \Bigl( \aE{6} + \bE{2} \epsilon + 
 {\cal O}(\epsilon^2) \Bigr) + {\cal O}(g^6) \Bigr], \hspace*{0.5cm}
 \label{mE} \\
 g_\rmi{E}^2 & = & T \Bigl[ g^2 + \frac{g^4}{(4\pi)^2}  
 \Bigl( \aE{7} +
 \bE{3} \epsilon + {\cal O}(\epsilon^2) \Bigr)  
 + {\cal O}(g^6) \Bigr] , 
 \label{gE} \\ 
 \lambda_\rmi{E}^{(1)} & = & T \Bigl[
 \frac{g^4}{(4\pi)^2} \Bigl( \bE{4} + 
 {\cal O}(\epsilon) \Bigr) + {\cal O}(g^6) \Bigr] , \\
 \lambda_\rmi{E}^{(2)} & = & T \Bigl[ 
 \frac{g^4}{(4\pi)^2}  
 \Bigl( \bE{5} + 
 {\cal O}(\epsilon) \Bigr) + {\cal O}(g^6) \Bigr] \;, \label{lE2} 
\ea
where $g^2$ is the renormalised coupling. 
We have named explicitly ($\aE{},\bE{}$) the
coefficients needed up to order ${\cal O}(g^6)$. 
The actual values for those needed at order ${\cal O}(g^6 \ln (1/g))$,
denoted by $\aE{}$,  are given in~Appendix~\ref{se:app}.  
The additional coefficients needed at the full order ${\cal O}(g^6)$ 
are denoted by $\bE{}$; some of these are also known
(for $\bE{4},\bE{5}$, e.g., see~\cite{quart}). 
The rest of the terms contribute only beyond ${\cal O}(g^6)$.

The expression for $p_\rmi{E}(T)$ is simply the functional 
integral in~\eq\nr{pqcd}, calculated to 4-loop level in the $\msbar$ 
scheme, but without any resummations. The only physical scale
entering is thus $2 \pi T$. The calculation has so far been carried 
out only to three loops~\cite{az,bn} so that $\bE{1}$ is not known. 
Even when performed with the fully renormalised theory, the results
in general contain uncancelled $1/\epsilon$ poles, as explicitly seen 
in the 3-loop expression in~\eq\nr{aE3} for $\aE{3}$. These only cancel 
when a physical fully resummed quantity is evaluated, i.e., 
in the sum $p_\rmi{QCD}=p_\rmi{E}+p_\rmi{M}+p_\rmi{G}$. Similarly, 
$m_\rmi{E}^2,g_\rmi{E}^2,\lambda_\rmi{E}^{(i)}$ can be obtained for
instance from suitable 2-, 3-, and 4-point functions, respectively.

%
\section{The complete result}

Combining now the results of Secs.~\ref{se:ggT}, \ref{se:gT}, \ref{se:2piT}
and expanding in $g$, we arrive at
\ba
 \frac{p_\rmi{QCD}(T)}{T^4 \mu^{-2 \epsilon} } \!\! & = & \!\! 
 \frac{p_\rmi{E}(T) + p_\rmi{M}(T) + p_\rmi{G}(T)}{T^4 \mu^{-2 \epsilon} } \nn
 \!\! & = & \!\!  
 g^0 \biggl\{ \aE{1} \biggr\} \nn 
 \!\! & + & \!\!  
 g^2 \biggl\{ \aE{2} \biggr\} \nn  
 \!\! & + & \!\!  
 \frac{g^3}{(4\pi)}
 \biggl\{ \frac{d_A}{3} \aE{4}^{3/2} \biggr\} \nn
 \!\! & + & \!\!   
 \frac{g^4}{(4\pi)^2} \biggl\{
 \aE{3} - d_A C_A
 \biggl[ 
 \aE{4} \biggl( 
 \frac{1}{4\epsilon} + \fr34 + \ln \frac{\bmu}{2 g T \aE{4}^{1/2}}
 \biggr)
 + \fr14 \aE{5} \biggr] \biggr\} \nn
 \!\! & + & \!\!  
 \frac{g^5}{(4\pi)^3} \biggl\{ d_A \aE{4}^{1/2} \biggl[
 \fr12 \aE{6} - C_A^2
 \biggl( 
 \frac{89}{24} + \frac{\pi^2}{6} - \frac{11}{6} \ln 2
 \biggr) 
 \biggr] \biggr\}  \nn
 \!\! & + & \!\!  
 \frac{g^6}{(4\pi)^4} \biggl\{
 \bE{1} - \fr14 {d_A} \aE{4} 
 \biggl[ 
 (d_A + 2) \bE{4} + 
 \frac{2 d_A - 1}{N_c} \bE{5} 
 \biggr] \nn
 & & \hspace*{0.7cm} 
 - d_A C_A \biggl[ 
 \fr14 \Bigl( \aE{6} + \aE{5} \aE{7} + 3 \aE{4} \aE{7} + \bE{2} 
 + \aE{4} \bE{3} \Bigr) \nn
 & & \hspace*{2.2cm}
 + \Bigl( \aE{6} + \aE{4} \aE{7} \Bigr) 
 \biggl(\frac{1}{4\epsilon} + \ln \frac{\bmu}{2 g T \aE{4}^{1/2}} \biggr)  
 \biggr] 
 \nn & & \hspace*{0.7cm}
 + d_A C_A^3 \biggl[ 
 \bM + \bG 
 + \aM \biggl( \frac{1}{\epsilon} + 8 \ln \frac{\bmu}{2 g T \aE{4}^{1/2}} 
 \biggr)
 +\aG \biggl( \frac{1}{\epsilon} + 8 \ln \frac{\bmu}{2 g^2 T C_A} \biggr)
 \biggr]
 \biggr\} \nn 
 \!\! & + & \!\! {\cal O}(g^7) + {\cal O}(\epsilon)\;. \label{full_result}
\ea
Utilising the expressions in~Appendix~\ref{se:app}, the terms up to 
order ${\cal O}(g^5)$ reproduce the known result in~\cite{zk}. 

For the contribution at order ${\cal O}(g^4)$, the 
$1/\epsilon$ divergence in $\aE{3}$ (cf.\ \eq\nr{aE3})
and the $1/\epsilon$ 
divergence from $p_\rmi{M}(T)$, shown explicitly in~\eq\nr{full_result}, 
cancel. This must happen
since $p_\rmi{QCD}(T)$ is a physical quantity. The associated 
$\bmu$'s also cancel, but a physical effect 
$\ln[m_\rmi{E}/(2\pi T)]\sim \ln(g\aE{4}^{1/2})$ remains~\cite{tt}. 

For the contribution at order ${\cal O}(g^6)$, a number of 
unknown coefficients remain (the $\bE{}$'s, $\bM$, $\bG$), but 
a similar cancellation is guaranteed to take place. In addition, 
the result must be  scale independent to the order it has been computed. 
The first point can be achieved by $\bE{1}$ (the other
$\bE{}$'s are finite), so that it has to have the structure
\be
 \bE{1} \equiv
 {d_A C_A} (\aE{6} + \aE{4} \aE{7} )
 \frac{1}{4\epsilon} 
 - {d_A C_A^3} (\aM + \aG)
 \frac{1}{\epsilon} 
 + \bE{6} \;, \label{bE5}
\ee
where $\bE{6}$ does not contain any $1/\epsilon$ poles. 
The latter point can be achieved by adding and subtracting 
$\ln [{\bmu}/({2 \pi T})]$'s, such that $\bmu$ gets effectively replaced 
by $2\pi T$ in the logarithms visible in the ${\cal O}(g^6)$ term
in~\eq\nr{full_result}. The $\ln [{\bmu}/({2 \pi T})]$'s left over, 
together with those coming from the $\bE{}$'s, 
serve to cancel the effects from the 2-loop running 
of $g^2(\bmu)$ and 1-loop running of $g^4(\bmu)$ in the lower order 
contributions, without introducing large logarithms. 

This general information is enough to fix the contributions
of order ${\cal O}(g^6 \ln (1/g))$ to $p_\rmi{QCD}(T)$. 
Indeed, after inserting~\eq\nr{bE5} and reorganising the 
logarithms appearing in the $\bE{}$'s as mentioned, 
there remains a logarithmic 4-loop term, 
\ba
 & & \hspace*{-1cm} 
 \left.
 \frac{p_\rmi{QCD}(T)}{T^4\mu^{-2 \epsilon}} 
 \right|_{g^6\ln(1/g)}
 \nn 
 & & \hspace*{-1cm} = 
 g^6 \frac{d_A C_A}{(4 \pi)^4}
 \biggl\{
 \Bigl( 
 \aE{6} + \aE{4} \aE{7} \Bigr) 
 \ln \Bigl(g \aE{4}^{1/2} \Bigr) 
 - 8 {C_A^2}
 \Bigl[ 
  \aM \ln \Bigl( g \aE{4}^{1/2} \Bigr) + 
  2 \aG \ln \Bigl(g C_A^{1/2} \Bigl)
 \Bigr] 
 \biggr\} \;, \label{main}
\ea
where $\aE{4}$ is in~\eq\nr{aE4}, 
$\aE{6}$ is in~\eq\nr{aE6}, 
$\aE{7}$ is in~\eq\nr{aE7}, 
$\aM$ is in~\eq\nr{aM}, and
$\aG$ is in~\eq\nr{aG}.  
Note that there are logarithms of
two types, with different non-analytic
dependences on group theory factors inside them.  
\eq\nr{main} is our main result. 

\vspace*{0.5cm}

Following~\cite{zk,bn}, let us finally insert $N_c = 3$,  and give 
also the numerical values for the various coefficients, for an 
arbitrary $N_f$. We obtain  
\ba
 p_\rmi{QCD}(T) & = & \frac{8\pi^2}{45} T^4 
 \biggl[ \sum_{i = 0}^6 p_i 
 \Bigl( \frac{\alpha_s(\bmu)}{\pi} \Bigr)^{i/2} \biggr] \;,  
 \label{pser}
\ea
where
\ba
 p_0 & = & 1 + \frac{21}{32} N_f \;, \\
 p_1 & = & 0 \;, \\
 p_2 & = & -\frac{15}{4}\Bigl( 1 + \frac{5}{12} N_f \Bigr) \;, \\
 p_3 & = & 30 \Bigl( 1 + \fr16 N_f \Bigr)^{3/2} \;, \\
 p_4 & = & 237.2 + 15.96 N_f - 0.4150 N_f^2
            + \frac{135}{2}\Bigl(1 + \fr16 N_f \Bigr)
              \ln\Bigl[\frac{\alpha_s}{\pi}
              \Bigl(1 + \fr16 N_f\Bigr)\Bigr] \nn 
     &   & -\frac{165}{8} \Bigl(1 + \fr5{12} N_f\Bigr) 
            \Bigl(1 - \frac{2}{33} N_f\Bigr) \ln \frac{\bmu}{2 \pi T} \;, \\
 p_5 & = & \Bigl(1 + \fr16 N_f\Bigr)^{1/2} \biggl[
           -799.1 - 21.96 N_f - 1.926 N_f^2 
    \nn 
     &   &  +\frac{495}{2} \Bigl(1 + \fr16 N_f\Bigr) 
            \Bigl(1 - \frac{2}{33} N_f\Bigr) 
            \ln \frac{\bmu}{2 \pi T}  \biggr] \;, \\
 p_6 & = &  \biggl[
            -659.2 - 65.89 N_f - 7.653 N_f^2 \nn 
     &   & + 
            \frac{1485}{2} \Bigl(1 + \fr16 N_f\Bigr) 
            \Bigl(1 - \frac{2}{33} N_f\Bigr) 
            \ln \frac{\bmu}{2 \pi T} \biggr] 
            \ln\Bigl[\frac{\alpha_s}{\pi}
            \Bigl(1 + \fr16 N_f\Bigr)\Bigr] \nn
     &   &  -475.6 \ln\frac{\alpha_s}{\pi} 
            + q_\rmi{a}(N_f)\, \ln^2 \frac{\bmu}{2 \pi T}
            + q_\rmi{b}(N_f)\, \ln \frac{\bmu}{2 \pi T}
            + q_\rmi{c}(N_f) \;, \label{p6}
\ea
where $q_\rmi{a}(N_f), q_\rmi{b}(N_f), q_\rmi{c}(N_f)$ are 
$\alpha_s$-independent polynomials in $N_f$. 
Two of them, $q_\rmi{a}(N_f)$, $q_\rmi{b}(N_f)$, can 
already be written down because they just cancel the 
$\bmu$-dependence arising from the terms of 
orders $\alpha_s(\bmu), \alpha_s^2(\bmu)$: 
\ba
 q_\rmi{a}(N_f) & = & -\frac{1815}{16} 
 \Bigl(1+\fr5{12} N_f \Bigr) \Bigl(1-\fr2{33} N_f\Bigr)^2
 \;, \\
 q_\rmi{b}(N_f) & = & 2932.9 +42.83 N_f -16.48 N_f^2 +0.2767 N_f^3
 \;.
\ea
The third one, $q_\rmi{c}(N_f)$, remains however unknown.

%
\section{The numerical convergence}
\label{se:numerics}

This Section is devoted to a numerical discussion of the result.
Since the $\mathcal{O}(g^6\ln(1/g))$ term cannot be given an unambiguous
numerical meaning until the $\mathcal{O}(g^6)$ term is specified, 
we have to present the result for various choices of the latter. 
In the relevant range of $T/\lambdamsbar$ the outcome will depend 
sensitively, even qualitatively, on this uncomputed term. One choice 
will be seen to agree with 4d lattice data down to about
$T/\lambdamsbar \sim 2...3$. Since however dimensional reduction, 
that is an effective description of QCD via the theory in~\eq\nr{eqcd}, 
is known to break down at about this point, and we have only kept
a finite number of terms in the expansion following from~\eq\nr{eqcd}, 
this cannot really be considered a prediction, even if the eventual
computation of the $\mathcal{O}(g^6)$ term gave just the appropriate value.
It is just an observation that a smooth transition from the domain of 
validity of our results to a domain of different approximations 
should be possible.

A standard procedure in the discussion of perturbative results would
be to take the expansion in~\eq\nr{pser} and to study whether
its scale dependence is reduced when further orders of perturbation
theory are included. As is well known since~\cite{az}, this fails
for the pressure, unless $T\gg \lambdamsbar$. Related to this, 
the numerical convergence of the perturbative expansion is known 
to be quite poor for any fixed scale choice, at least
for temperatures below the electroweak scale~\cite{zk,bn,adjoint}.
The new term we have computed does not change this general pattern.
But the culprit is known: it is $p_\rmi{M}(T) + p_\rmi{G}(T)$
emerging from the 3d sector of the theory,
where the expansion parameter is only
$g_\rmi{E}^2/(\pi m_\rmi{E})\sim g/\pi$. In contrast, for
$p_\rmi{E}(T)$ as well as for, say, jet physics, the expansion
parameter is $\alpha_s/\pi$, and there are good reasons to
expect numerical convergence to be much better.

For these reasons, we will only discuss the sensitivity of 
the result on the so far unknown $\mathcal{O}(g^6)$ coefficient,
as well as the slow convergence of the 3d sector, in the following. 
For simplicity, we only consider the case $N_c=3, N_f=0$ here.

As in~\cite{a0cond}, the actual form we choose for plotting 
contains $p_\rmi{M}(T) + p_\rmi{G}(T)$  (\eqs\nr{gT_contr} +  
\nr{resG}) in an ``un-expanded'' form, that is, 
with $m_\rmi{E}, g_\rmi{E}^2$ inserted from~\eqs\nr{mE}, \nr{gE}, 
and $g_\rmi{M}^2$ from \eq\nr{gM}. 
This means that we are effectively summing up higher orders: 
the ${\cal O}(g^3)$-term is really 
${\cal O}(g^2 + g^4)^{3/2}$, 
while the ${\cal O}(g^6\ln(1/g))$ term contains a resummed
coefficient, being then effectively ${\cal O}((g^2+ g^4)^3 \ln(1/g))$. 
We proceed in this way because then a comparison with numerical 
determinations~\cite{a0cond} of the slowly convergent 
part $p_\rmi{M}(T)+p_\rmi{G}(T)$ 
is more straightforward, 
and also because the resummations carried out reduce the 
$\bmu$-dependence of the outcome. However, we have checked 
that the practical conclusions remain the same even if we 
plot directly the expression in~\eqs\nr{pser}--\nr{p6}
(but with a larger scale dependence). 

\begin{figure}[tb]

\centerline{
    \epsfxsize=7.5cm\epsfbox{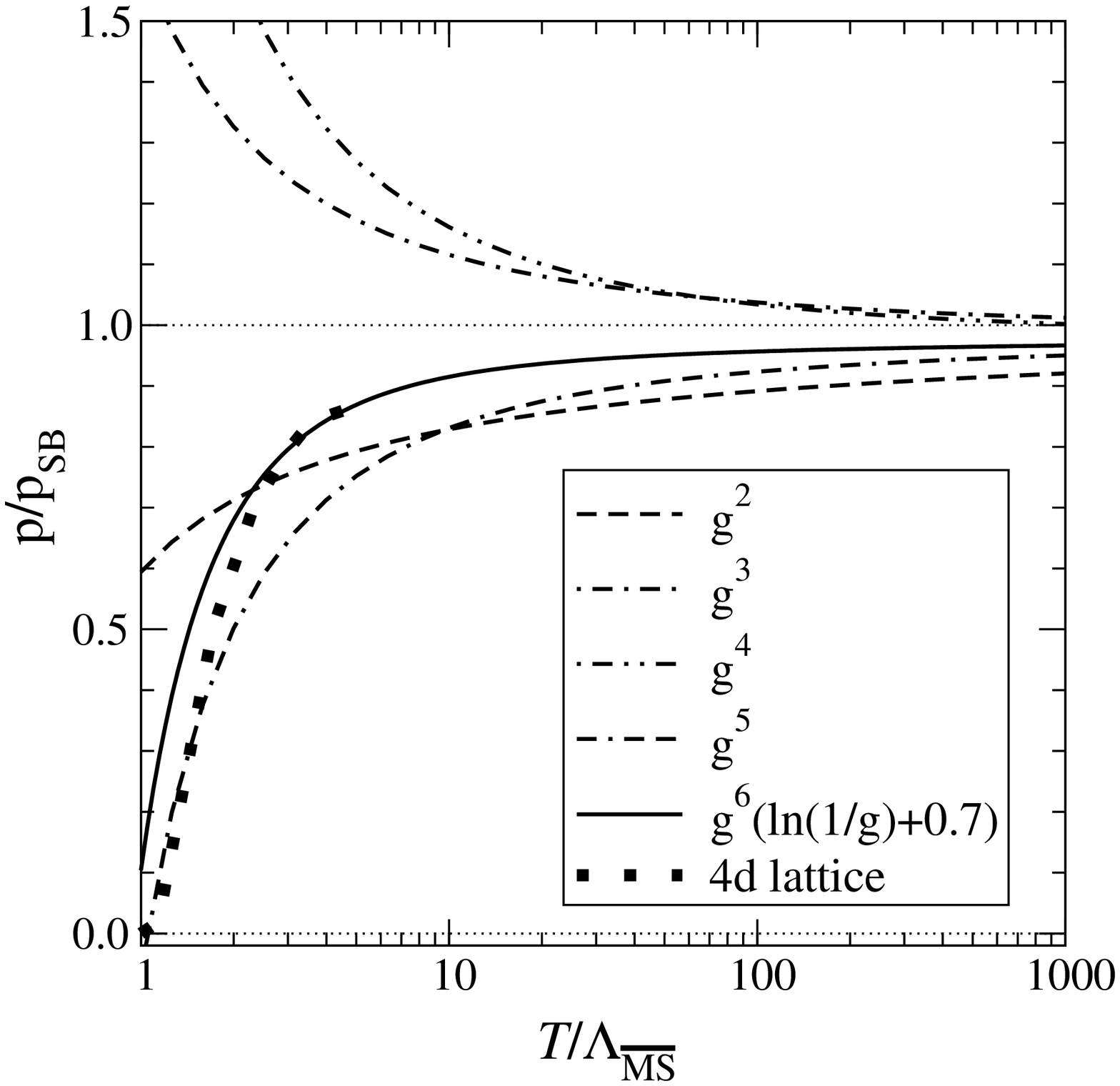}%
    \epsfxsize=7.5cm\epsfbox{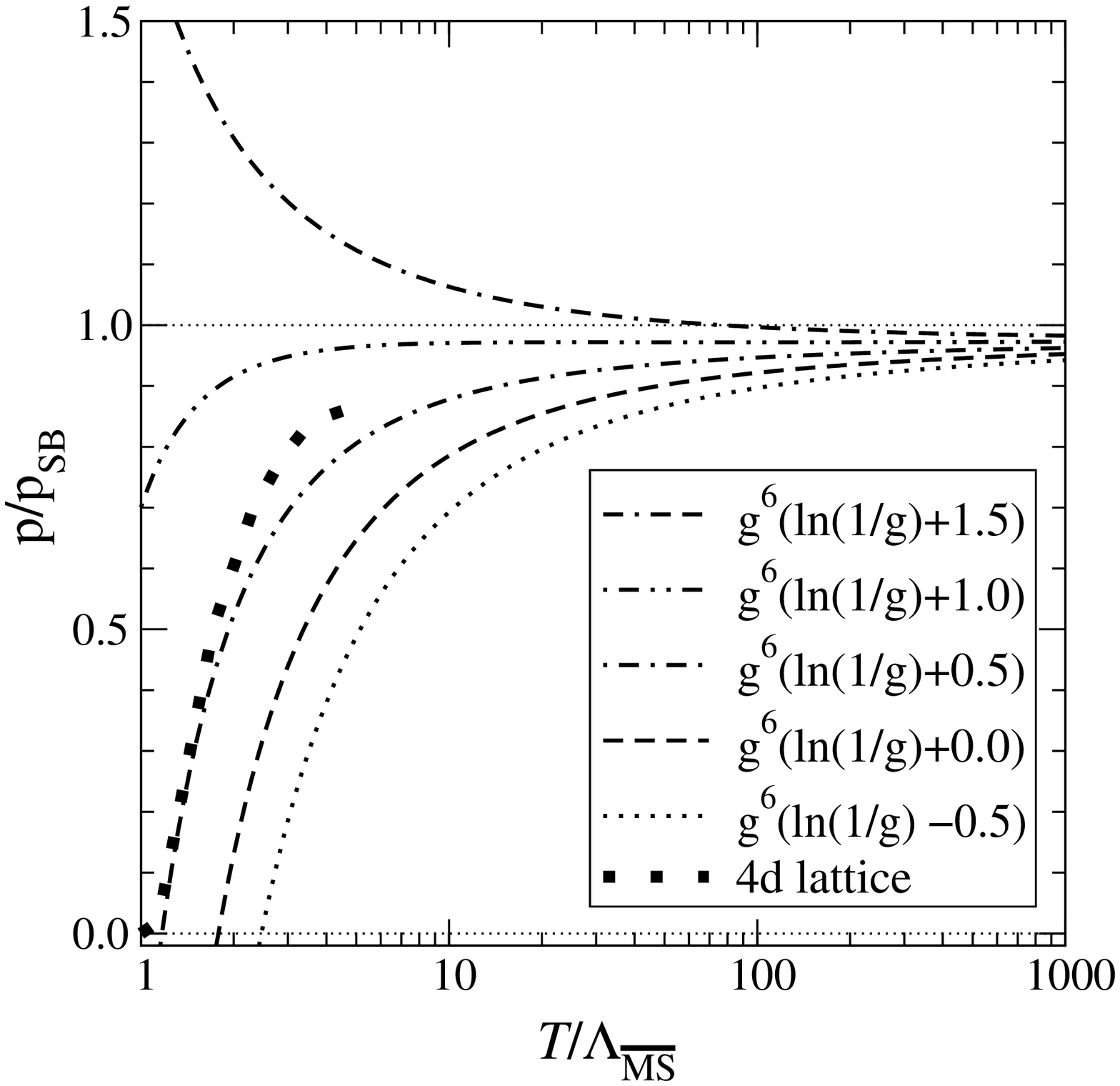}}

\vspace*{0.5cm}

\caption[a]{Left: perturbative results at various orders (the precise
meanings thereof are explained in~\se\ref{se:numerics}), including
${\cal O}(g^6)$ for an optimal constant, normalised to the 
non-interacting Stefan-Boltzmann value $p_\rmi{SB}$. 
Right: the dependence of the 
${\cal O}(g^6)$ result on the (not yet computed) constant, which
contains both perturbative and non-perturbative contributions.
The 4d lattice results are from~\cite{boyd}.}

\label{fig:pert_g6}
\end{figure}

To be specific, the genuine ${\cal O}(g^6\ln(1/g) + g^6)$ contribution, 
which collects the effects from all the terms involving the
$\bE{}$'s, $\bM$, $\bG$, $\aM$, and $\aG$ in~\eq\nr{full_result}, 
is now written in the form (specific for $N_c=3,N_f=0$, 
where $m_\rmi{E}/g_\rmi{E}^2 \sim 1/g $), 
\be
  \delta \left[ \frac{p_\rmi{QCD}(T)}{T \mu^{-2 \epsilon}}
  \right]_{g^6\ln(1/g)} \, \equiv \, 
  8 d_A C_A^3 \frac{g_\rmi{E}^6}{(4\pi)^4} 
  \Bigl[(\aM + 2 \aG)\ln \frac{m_\rmi{E}}{g_\rmi{E}^2} + \delta \Bigr]\;,
\ee
while the remaining ${\cal O}(g^6)$ terms of~\eq\nr{full_result} 
are contained in the resummed lower order contributions. 
The results are shown in~\fig\ref{fig:pert_g6}, for various
values of $\delta$. The power of $g$ labelling the curves indicates 
the leading magnitude of each resummed contribution. The scale
is chosen as $\bmu \approx 6.7 T$, as suggested by the 
next-to-leading order expression for $g_\rmi{E}^2$~\cite{adjoint}.
We observe that for a specific value of $\delta$, 
the curve extrapolates well to 4d lattice data. 

While \fig\ref{fig:pert_g6} looks tempting, the question still remains whether 
the good match to 4d lattice data with a specific value 
of the constant is simply a coincidence. This issue can be fully  
settled only once the constant is actually computed. However, we
can already inspect how the slowly convergent part of the 
pressure, $p_\rmi{M}+p_\rmi{G}$, really behaves.

\begin{figure}[tb]

\centerline{
    \epsfxsize=7.5cm\epsfbox{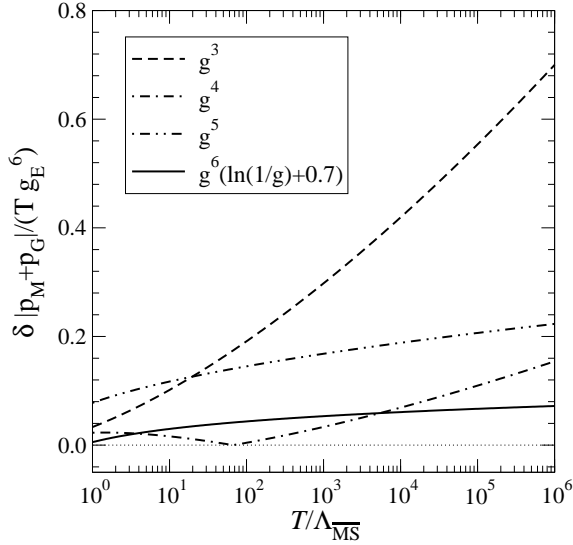}}

\vspace*{0.5cm}

\caption[a]{
The absolute values of the various terms of the slowly convergent
expansion for $p_\rmi{M}(T)+p_\rmi{G}(T)$, 
normalised by $T g_\rmi{E}^6$.}

\label{fig:pM}
\end{figure}

The different finite
terms in $(p_\rmi{M}+p_\rmi{G})/(T g_\rmi{E}^6)$ 
are plotted in Fig.\ref{fig:pM}.
The $\lambda_\rmi{E}^{(i)}$-contributions
are negligible. The results depend then essentially
only on $m_\rmi{E}^2/g_\rmi{E}^4$, which for $N_c=3,N_f=0$  
is $m_\rmi{E}^2/g_\rmi{E}^4 \approx 0.32 \log_\rmi{10} (T/\lambdamsbar)  
+0.29$. 
We observe that the leading 1-loop term ${\cal O}(g^3)$ is dominant
for $T/\lambdamsbar\gsim 10$,
the 3-loop term ${\cal O}(g^5)$ is rather big, bigger in absolute value 
than the 2-loop term ${\cal O}(g^4)$ within the $T$-range of the figure, while
the 4-loop term is always very small. Therefore, while it is well possible
that there is again a big ``odd'' ${\cal O}(g^7)$ contribution, 
it is perhaps not completely outrageous either 
to hope that the convergence could also already be reasonable, 
once the full ${\cal O}(g^6)$ contribution is included. If this were
the case, then all higher order contributions would have to sum up
to a small number. 

Finally, it is perhaps interesting to remark that at the 
time of the numerical lattice Monte Carlo 
study in ref.~\cite{a0cond}, nothing was known about
the coefficient $\bE{1}$, which was therefore set to zero
(cf.\ \eq(4) in~\cite{a0cond}), while 
the part $p_\rmi{M}(T)+p_\rmi{G}(T)$ was determined non-perturbatively. 
But this means that a logarithmic term coming from the scale $2\pi T$, 
$\sim -g^6 (\aM + \aG) \ln [\bmu/(2\pi T)]$, was missed. With the 
scale choice $\bmu\equiv \bmu_\rmi{E} = g_\rmi{E}^2$
within results obtained with ${\cal L}_\rmi{E}$, 
this converted to a missing 
${\cal O}(g^6 \ln(1/g))$ contribution
$g^6 (2\aM+2\aG)\ln(1/g)$. With the same scale choice
the non-perturbative part, on the other hand, 
contributed $ -g^6\aM\ln(1/g)$ and led to the wrong
curvature of the pressure seen at small $T/\lambdamsbar$.
Adding the missing part, which now has been computed, leads to a 
total of $g^6(\aM+2\aG)\ln(1/g)$ with the opposite sign and the correct
({\it i.e.}, the one seen in 4d lattice measurements)
curvature in~\fig\ref{fig:pert_g6} (for small values of $\delta$).
Therefore the ${\cal O}(g^6 \ln(1/g))$ terms are indeed physically
very relevant.

%
\section{Conclusions}
\label{se:conclusions}

We have addressed in this paper  the 4-loop logarithmic contributions
to the pressure of hot QCD. Physical (regularisation independent) 
logarithms can only arise from a ratio of two scales. Since there 
are three parametrically different scales in the system, $2\pi T, gT, g^2T$, 
there are then various types of perturbatively 
computable logarithms in the 4-loop expression for the pressure:
\begin{itemize}
\item[1.] 
Logarithms of the type $g^6 \ln[(2\pi T)/(g^2 T)]$. The coefficient
of these is computed in~\cite{ys}, and given in~\eq\nr{aG}.  
\item[2.]
Logarithms of the type $g^6 \ln[(2\pi T)/(g T)]$. The coefficient 
of these is computed in~\cite{aminusb}, and given in~\eq\nr{aM}. 
\item[3.]
Logarithms related to the running of the coupling constant in 
the 3-loop expression of order ${\cal O}(g^4 \ln[(2\pi T)/(g T)])$. 
Their $\msbar$
coefficient can be seen in the first term in~\eq\nr{main}, but it 
depends on the scheme, and can in principle even be chosen to vanish.
\end{itemize}
Logarithms of the first and second types can be
written in many ways: it may be more intuitive, for instance, 
to reorganise them as 
\be
 g^6  \aG \ln \biggl( \frac{2\pi T}{g^2 T} \biggr)
  + g^6 \aM \ln \biggl( \frac{2\pi T}{gT} \biggr)  = 
 g^6 (\aM + \aG) \ln \biggl( \frac{2\pi T}{gT} \biggr) +  
 g^6 \aG \ln \biggl( \frac{g T}{g^2 T} \biggr) \;.
\ee

The existence of three kinds of logarithms is somewhat specific to non-Abelian
gauge theory. In QED, in particular, none of the logarithms appear.
This is due to the fact that the effective theories we have used for their
computation, \eqs\nr{eqcd}, \nr{mqcd}, are non-interacting
(apart from a term $\sim A_0^4$ in~\eq\nr{eqcd}, which does
not lead to logarithms). Therefore we have nothing to add to the known
${\cal O}(g^5)$ QED result obtained in~\cite{qed}. 
In the $\phi^4$ scalar theory, on 
the other hand, there is a logarithm of the second type, and also one 
somewhat analogous to the third type. Their coefficients were already  
computed in~\cite{phi4}.

There are interesting checks that can be made on the  
various logarithms mentioned, using methods completely 
different from those employed here. For instance, logarithms of 
the first and second types could in principle be seen with 
3d lattice Monte Carlo methods~\cite{klpr,latt}, as well 
as with stochastic perturbation theory~\cite{fdr}.
A very interesting analytical check would be to compute 
the 4-loop free energy directly in 4d in strict dimensional 
regularisation, but without any resummation. By definition, this computation 
produces the coefficient $\bE{1}$ in~\eq\nr{pE}~\cite{bn}, and
one check is that the result must contain the $1/\epsilon$
divergences shown in~\eq\nr{bE5}.  

To complete the free energy from 
the current level ${\cal O}(g^6\ln(1/g))$ 
to the full level ${\cal O}(g^6)$, 
would require significantly more work than the computation presented 
here. More specifically, there are contributions from all the 
scales in the problem, ranging from $2 \pi T$ 
(the coefficients $\bE{1},...,\bE{5}$), through $gT$
(the coefficient $\bM$), down to the
non-perturbative scale $g^2T$ 
(the coefficient $\bG$). This then requires carrying out 
4-loop finite temperature sum-integrals, 4-loop vacuum integrals
in $d=3-2\epsilon$, 4-loop vacuum integrals in 3d
lattice regularisation, and lattice simulations of the pure 3d
gauge theory in~\eq\nr{mqcd}. 

Nevertheless, given the potentially important combined effect
of all these contributions, as indicated by~\fig\ref{fig:pert_g6}, 
such computations would clearly be well motivated.

%
\begin{acknowledgments}
This work was partly supported by the RTN network {\em Supersymmetry and 
the Early Universe}, EU Contract no.\ HPRN-CT-2000-00152, by the Academy
of Finland, Contracts no.\ 77744 and 80170, by the 
DOE, under Cooperative Agreement no.~DF-FC02-94ER40818, and 
by the National Science Foundation, under Grant no.\ PHY99-07949.
We thank KITP, Santa Barbara, where part of this work was
carried out, for hospitality.  
\end{acknowledgments}


\appendix
\renewcommand{\theequation}{\Alph{section}.\arabic{equation}}


\section{Matching coefficients}
\label{se:app}

In \eqs\nr{pE}--\nr{lE2} we have defined a number of matching 
coefficients, the $\aE{}$'s and $\bE{}$'s. For the $\aE{}$'s, 
the following expressions can be extracted from~\cite{hl,generic,bn}: 
\ba
 \aE{1} & = &   
 \frac{\pi^2}{180} \Bigl( 4 d_A + 7 d_F \Bigr) 
 \;, \label{aE1} \\ 
 \aE{2} & = &   
 -\frac{d_A}{144}  \Bigl(C_A + \fr52 T_F \Bigr)
 \;, \label{aE2} \\ 
 \aE{3} & = &   
 \frac{d_A}{144} 
 \biggl[ 
 C_A^2 \biggl( 
 \frac{12}{\epsilon} + \frac{194}{3} \logT + \frac{116}{5} + 4 \gamma
 + \frac{220}{3}\logz{1} - \frac{38}{3}\logz{3}
 \biggr) \nn 
 & + & 
 C_A T_F \biggl(  
 \frac{12}{\epsilon} + \frac{169}{3} \logT + \frac{1121}{60} 
 - \frac{157}{5} \ln 2 + 8 \gamma + \frac{146}{3} \logz{1} - 
 \fr13 \logz{3}
 \biggr) \nn 
 & + & 
 T_F^2 \biggl( 
 \frac{20}{3} \logT + \fr13 - \frac{88}{5} \ln 2 + 4 \gamma
 + \frac{16}{3} \logz{1} - \fr83 \logz{3}
 \biggr) \nn 
 & + & 
 C_F T_F \biggl( 
 \frac{105}{4} - 24 \ln 2 
 \biggr)  
 \biggr] 
 \;,  \label{aE3} \\ 
 \aE{4} & = &  
 \fr13 (C_A + T_F) 
 \;, \label{aE4} \\ 
 \aE{5} & = &   
 \fr23 \biggl[ 
 C_A \biggl( 
 \ln\frac{\bmu}{4 \pi T}
 + \frac{\zeta'(-1)}{\zeta(-1)} 
 \biggr)
 + T_F \biggl( \ln\frac{\bmu}{4 \pi T}
 + \fr12 - \ln 2 + \frac{\zeta'(-1)}{\zeta(-1)} 
 \biggr)
 \biggr] 
 \;, \label{aE5} \\ 
 \aE{6} & = &   
 C_A^2 \biggl(\frac{22}{9} \ln\frac{\bmu e^\gamma}{4 \pi T} + \fr59  \biggr) 
 + C_A T_F \biggl(
  \frac{14}{9} \ln\frac{\bmu e^\gamma}{4 \pi T}
  -\frac{16}{9} \ln 2
  +   1 \biggr) \nn 
 & + & 
 T_F^2 \biggl(- \frac{8}{9} \ln\frac{\bmu e^\gamma}{4 \pi T} 
 -\frac{16}{9} \ln 2 
 + \fr49  \biggr) 
 - 2 C_F T_F 
 \;, \label{aE6} \\ 
 \aE{7} & = &   
 C_A \biggl(
 \frac{22}{3}   \ln\frac{\bmu e^\gamma}{4 \pi T} 
 + \fr13 \biggr)
 - T_F \biggl( \fr83  \ln\frac{\bmu e^\gamma}{4 \pi T} 
 +\frac{16}{3} \ln 2 \biggr)
 \;. \label{aE7} 
\ea
Note that with our notation, the 1-loop running of the renormalised 
coupling constant goes as
\be
 g^2(\bmu) = g^2(\bmu_0) - \fr23 (11 C_A - 4 T_F) 
 \frac{g^4(\bmu_0)}{(4 \pi)^2} \ln\frac{\bmu}{\bmu_0}\;.\label{gmu}
\ee



\begin{thebibliography}{99}

\bibitem{linde}
A.D.~Linde,
Phys.\ Lett.\ {\bf B96}, 289 (1980).

\bibitem{gpy}
D.J.~Gross, R.D.~Pisarski and L.G.~Yaffe,
Rev.\ Mod.\ Phys.\ {\bf 53}, 43 (1981).

\bibitem{es}
E.V.~Shuryak,
Sov.\ Phys.\ JETP {\bf 47}, 212 (1978)
[Zh.\ Eksp.\ Teor.\ Fiz.\  {\bf 74}, 408 (1978)];
S.A.~Chin,
Phys.\ Lett.\ {\bf B78}, 552 (1978).


\bibitem{jk}
J.I.~Kapusta,
Nucl.\ Phys.\ {\bf B148}, 461 (1979).

\bibitem{tt}
T.~Toimela,
Phys.\ Lett.\ {\bf B124}, 407 (1983).

\bibitem{az}
P.~Arnold and C.~Zhai,
Phys.\ Rev.\  {\bf D50}, 7603 (1994)
\eprint{[hep-ph/9408276]};
%
{\it ibid.}\  {\bf D51}, 1906 (1995)
\eprint{[hep-ph/9410360]}.

\bibitem{zk}
C.~Zhai and B.~Kastening,
Phys.\ Rev.\  {\bf D52}, 7232  (1995)
\eprint{[hep-ph/9507380]}.

\bibitem{gdm}
G.D.~Moore,
JHEP {\bf 0210}, 055 (2002)
\eprint{[hep-ph/0209190]};
%
A.~Ipp, G.D.~Moore and A.~Rebhan,
\eprint{hep-ph/0301057}.

\bibitem{ys}
Y.~Schr\"oder, 
{\it Logarithmic divergence in the energy density
of the three-dimensional Yang--Mills theory},
in preparation. 

\bibitem{aminusb}
K.~Kajantie, M.~Laine, K.~Rummukainen and Y.~Schr\"oder, 
{\it Four-loop vacuum energy density of
the SU($N_c$) + adjoint Higgs theory},
in preparation. 

\bibitem{bn}
E. Braaten and A. Nieto,
Phys.\ Rev.\ {\bf D53}, 3421  (1996)
\eprint{[hep-ph/9510408]}.

\bibitem{adjoint}
K.~Kajantie, M.~Laine, K.~Rummukainen and M.~Shaposhnikov,
Nucl.\ Phys.\ {\bf B503}, 357 (1997)
\eprint{[hep-ph/9704416]}.

\bibitem{dr}
P. Ginsparg, 
Nucl.\ Phys.\ {\bf B170}, 388 (1980);
%
T. Appelquist and R.D. Pisarski,
Phys.\ Rev.\ {\bf D23}, 2305 (1981).

\bibitem{a0cond}
K.~Kajantie, M.~Laine, K.~Rummukainen and Y.~Schr\"oder,
Phys.\ Rev.\ Lett.\  {\bf 86}, 10 (2001)
\eprint{[hep-ph/0007109]}.

\bibitem{generic}
K.~Kajantie, M.~Laine, K.~Rummukainen and M.~Shaposhnikov,
Nucl.\ Phys.\ {\bf B458}, 90  (1996)
\eprint{[hep-ph/9508379]}.

\bibitem{sc}
S.~Chapman,
Phys.\ Rev.\ {\bf D50}, 5308 (1994)
\eprint{[hep-ph/9407313]}.

\bibitem{ys_proc}
Y.~Schr\"oder, 
\eprint{hep-ph/0211288}.

\bibitem{jamv}
J.A.M.~Vermaseren,
\eprint{math-ph/0010025};
{\tt http://www.nikhef.nl/\~{}form/}.

\bibitem{sd}
K.~Kajantie, M.~Laine and Y.~Schr\"oder,
Phys.\ Rev.\ {\bf D65}, 045008  (2002)
\eprint{[hep-ph/0109100]}.

\bibitem{fkrs}
K.~Farakos, K.~Kajantie, K.~Rummukainen and M.E.~Shaposhnikov,
Nucl.\ Phys.\ {\bf B425}, 67 (1994)
\eprint{[hep-ph/9404201]}.

\bibitem{quart}
S.~Nadkarni,
Phys.\ Rev.\ {\bf D38}, 3287 (1988);
%
N.P.~Landsman,
Nucl.\ Phys.\ {\bf B322}, 498 (1989).

\bibitem{boyd}
G.~Boyd {\it et al.}, 
Nucl.\ Phys.\  {\bf B469}, 419  (1996)
\eprint{[hep-lat/9602007]};
%
A.~Papa,
Nucl.\ Phys.\  {\bf B478}, 335  (1996)
\eprint{[hep-lat/9605004]};
%
B.~Beinlich, F.~Karsch, E.~Laermann and A.~Peikert,
Eur.\ Phys.\ J.\  {\bf C6}, 133  (1999)
\eprint{[hep-lat/9707023]};
%
M.~Okamoto {\it et al.}  [CP-PACS Collaboration],
Phys.\ Rev.\  {\bf D60}, 094510  (1999)
\eprint{[hep-lat/9905005]}.

\bibitem{qed}
R.R.~Parwani,
Phys.\ Lett.\ {\bf B334}, 420 (1994); 
{\it ibid.}\  {\bf B342}, 454 (1995) (E)
\eprint{[hep-ph/9406318]};
%
R.R.~Parwani and C.~Corian\`o,
Nucl.\ Phys.\ {\bf B434}, 56 (1995)
\eprint{[hep-ph/9409269]}.

\bibitem{phi4}
E.~Braaten and A.~Nieto,
Phys.\ Rev.\ {\bf D51}, 6990 (1995)
\eprint{[hep-ph/9501375]}.

\bibitem{klpr}
F.~Karsch, M.~L\"utgemeier, A.~Patk\'os and J.~Rank,
Phys.\ Lett.\ {\bf B390}, 275 (1997)
\eprint{[hep-lat/9605031]}.

\bibitem{latt}
K.~Kajantie, M.~Laine, K.~Rummukainen and Y.~Schr\"oder,
Nucl.\ Phys.\ (Proc.\ Suppl.)\  {\bf B106}, 525 (2002)
\eprint{[hep-lat/0110122]};
%
\eprint{hep-lat/0209072}.

\bibitem{fdr}
F.~Di Renzo, E.~Onofri, G.~Marchesini and P.~Marenzoni,
Nucl.\ Phys.\ {\bf B426}, 675 (1994)
\eprint{[hep-lat/9405019]}.

\bibitem{hl}
S.~Huang and M.~Lissia,
Nucl.\ Phys.\ {\bf B438}, 54 (1995)
\eprint{[hep-ph/9411293]}.

\end{thebibliography}
\end{document}